# *In vivo* Optical Coherence Elastography Reveals Spatial Variation and Anisotropy of Corneal Stiffness

Guo-Yang Li, Xu Feng, and *Seok-Hyun Yun

*Abstract*— *Objective*: The mechanical properties of corneal tissues play a crucial role in determining corneal shape and have significant implications in vision care. This study aimed to address the challenge of obtaining accurate *in vivo* data for the human cornea. *Methods:* We have developed a high-frequency optical coherence elastography (OCE) technique using shear-like antisymmetric (A0)-mode Lamb waves at frequencies above 10 kHz. *Results:* By incorporating an anisotropic, nonlinear constitutive model and utilizing the acoustoelastic theory, we gained quantitative insights into the influence of corneal tension on wave speeds and elastic moduli. Our study revealed significant spatial variations in the shear modulus of the corneal stroma on healthy subjects for the first time. The central cornea exhibited a shear modulus of 74 kPa, while the corneal periphery showed a decrease to 41 kPa. The limbus demonstrated an increased shear modulus exceeding 100 kPa. We obtained wave displacement profiles that are consistent with highly anisotropic corneal tissues. *Conclusion:* Our approach enabled precise measurement of corneal tissue elastic moduli *in situ* with high precision (< 7%) and high spatial resolution (< 1 mm). *Significance:* The high-frequency OCE technique holds promise for biomechanical evaluation in clinical settings, providing valuable information for refractive surgeries, degenerative disorder diagnoses, and intraocular pressure assessments.

## I. Introduction

The mechanical properties of the cornea play a crucial role in determining its response to mechanical stress and have significant implications in vision care. Achieving optimal refractive outcomes in refractive surgeries and accurately measuring intraocular pressure (IOP) in tonometry relies on understanding corneal mechanics while considering corneal stiffness variability. Corneal protrusion in keratoconus (KC) [1, 2], a degenerative disorder, serves as a potential diagnostic marker for localized mechanical degradation [3, 4]. Corneal crosslinking (CXL), a treatment for KC and corneal ectasia, aims to increase stromal elastic modulus and regenerate collagen fibers. Therefore, the ability to measure corneal tissue stiffness is highly valuable in these medical procedures.

While various mechanical tools exist for characterizing corneal tissues *ex vivo* [5-7], *in vivo* measurements pose significant challenges. Some promising approaches include commercial instruments like the Ocular Response Analyzer and Corvis ST provide overall corneal stiffness indices without spatial resolution [8, 9]. Brillouin microscopy maps longitudinal elastic modulus with high resolution [10, 11] but lacks shear and tensile moduli measurements required for describing corneal deformation.

Optical coherence elastography (OCE) is an emerging technology for corneal characterization [12-18]. By exciting elastic waves in the cornea and measuring their propagation speeds, OCE allows direct calculation of shear or tensile elastic modulus. While noncontact methods to excite elastic waves are appealing, options such as air puff, shear ultrasound waves, and acoustic radiation force have not yet achieved the desired efficiency for exciting elastic waves at high frequencies (> 10 kHz) that are crucial for achieving high accuracy and resolution [19]. Our previous work conducted the first-in-human OCE measurement and obtained shear moduli in the central cornea with wave frequency up to 16 kHz [16].

In this study, we present a more detailed investigation of the human cornea using advanced wave analysis. By applying the acoustoelastic theory to a corneal model accounting for IOP-induced tension, nonlinearities, anisotropy, and spatial variations of tissue stiffness, we have obtained new *in vivo* data. Our findings reveal a lower shear modulus in the peripheral cornea compared to the central cornea. Additionally, we quantitatively derive the ratio of tensile to shear moduli to reveal the mechanical anisotropy. This work represents a substantial advancement of a preliminary version reported in ref [20].

## II. Theory and Methods

### A. Optical coherence elastography system

We utilized a custom-built optical coherence tomography (OCT) system equipped with a swept laser source centered at 1300 nm [16, 21]. The system operated at an A-line rate of 43.2 kHz, delivering an optical power of 10 mW to the cornea.

Manuscript received XXX; revised XXX; accepted XXX. Date of publication XXX; date of current version XXX.

Guo-Yang Li and Xu Feng contribute equally to this work. Corresponding author: Seok-Hyun Yun.

Guo-Yang Li, Xu Feng, and Seok-Hyun Yun are with Harvard Medical School and Wellman Center for Photomedicine, Massachusetts General Hospital. Guo-Yang Li is currently with the Department of Mechanics and Engineering Science, College of Engineering, Peking University.

Seok-Hyun Yun is with the Harvard-MIT Division of Health Sciences and Technology. (correspondence e-mail: syun@hms.harvard.edu).

This study is supported by the U.S. National Institutes of Health (NIH) via grants R01EB027653, R01EY033356, and R01EY034857.

Digital Object Identifier XXX.

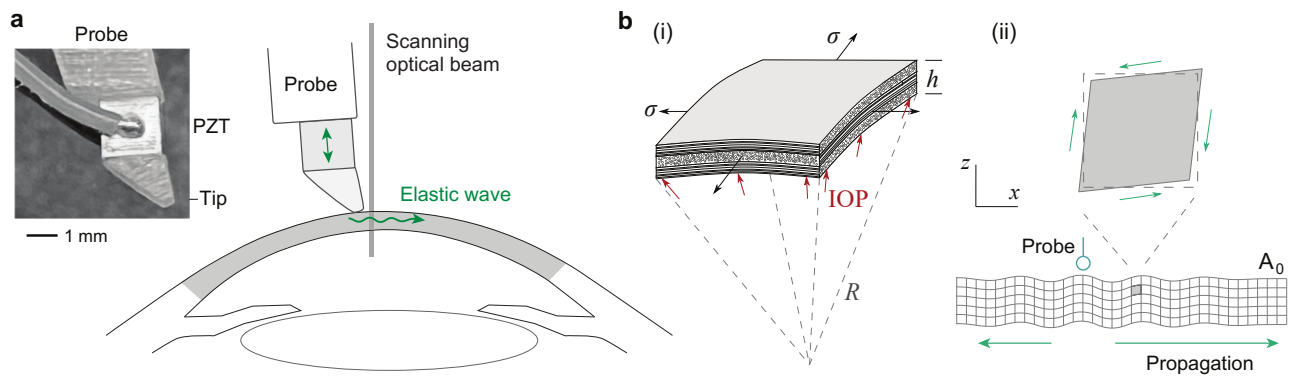

Figure 1. (a) Schematic of optical coherence elastography using a contact probe. The inset shows a photo of the probe consisting of a piezoelectric (PZT) transducer and a tip. (b) Biomechanical model of the cornea and an excited elastic wave. Schematic (i) illustrates the microstructure of the stroma under intraocular pressure (IOP) and in-plane tension σ. Schematic (ii) demonstrates the A0 mode wave along the cornea, involving shear deformation.

Elastic waves were generated using a contact probe, covering a broad frequency range from 2 to 30 kHz. Figure 1a presents a schematic of the experimental setup. The probe consisted of a piezoelectric transducer (PZT) and a tip with a radius of curvature of 0.4 mm. A gentle contact force of approximately 0.01 N was applied as the tip contacted the corneal surface.

At each location of the OCT beam, we acquired 360 A-lines (M-scan). The acquired M-scan data underwent Fourier transformation to extract the wave amplitudes and phases along the depth at the respective beam location. Depending on the wave frequency, data were collected at 96 transverse points along a distance ranging from 2 to 10 mm. The wave phase velocity at each frequency $f$ was determined by performing a Fourier transform of the complex wave amplitude data from the 96 points, allowing us to identify the peak wavenumber $k$, or by deriving the gradient of the wave phase that should equal to $k$. The phase velocity ($v$) was then calculated using the equation $v = 2\pi f/k$. The acquired dataset was used to generate a wave image, representing a snapshot of the wave at a specific phase of oscillation.

*B. The microstructure and mechanical stress of the cornea*

The corneal stroma is composed of lamellae, consisting of finely arranged collagen fibrils along the plane of the cornea [22, 23]. The anterior cornea exhibits more interwoven and undulating collagen bundles, while the mid and posterior corneas display orthogonally arranged lamellae [24]. This unique microstructure gives the cornea its distinctive elastic properties [25, 26]. The cornea is commonly modeled as an anisotropic composite plate (Fig. 1b (i)), with mechanically reinforcing fibers aligned along the plane [27, 28]. These fibers enhance the tissue's stiffness along the plane. The collagen fibers contribute to nonlinear elasticity or hyper-elasticity, where stiffness increases with strain. However, their influence on the shear modulus is thought to be relatively minimal and independent of the shearing direction. The corneal tissue exhibits significant mechanical anisotropy due to the pronounced differences between in-plane and out-of-plane tensile moduli, as well as between in-plane tensile and shear moduli.

While the compressional stress from IOP exerted on the posterior surface has minimal impact on tensile and shear moduli, IOP induces in-plane tension in the corneal tissue. This tension significantly increases the tensile modulus by stretching the collagen fibers. According to the Young-Laplace equation, the tensional stress σ parallel to the plane is given by

$$\sigma = IOP * (R/2h) \quad (1)$$

Here, $R$ and $h$ represent the radius of curvature and thickness of the cornea, respectively. With $R = 7.8$ mm and $h = 550$ μm, we find that $\sigma \approx 7.1 * IOP$. For $IOP = 15$ mmHg = 2.0 kPa, $\sigma \approx 14.2$ kPa. Such an in-plane tension significantly alters the in-plane tensile modulus of the cornea. Additionally, corneal tension affects the speed of elastic waves utilized in OCE through the acoustoelastic effect, which we will describe in the next section.

*C. The acoustoelastic theory*

The cornea is a waveguide for elastic waves due to its surrounding air and aqueous humor [16, 29-31]. With our contact vibrational probe, we primarily excite and analyze the fundamental antisymmetric Lamb wave known as the A0 mode. Figure 1b(ii) illustrates the flexural deformation profile of the wave. In a tension-free isotropic material, the wave speed of the A0 mode would be determined by the shear modulus of the material [32]. However, in actual corneal tissues, which are anisotropic, nonlinear, and under tension, the wave speed depends on both shear and in-plane tensile moduli, as well as the tension σ.

According to the acoustoelastic theory [33, 34], the propagation of a plan elastic wave can be described using three parameters, $\alpha$, $\beta$, and $\gamma$, which characterize material stiffness and stress of the cornea. It can be shown that (see Appendix A and B)

$$\sigma = \alpha - \gamma \quad (2)$$

The stress σ explicitly affects the propagation of elastic waves. The parameters are related to the out-of-plane shear modulus $\bar{G}_{zx}$ and in-plane tensile modulus $\bar{E}_{xx}$ of the prestressed cornea, via (see Appendix C)

$$\bar{G}_{zx} \triangleq \alpha \text{ and } \bar{E}_{xx} \triangleq 2\beta + 2\gamma \quad (3)$$

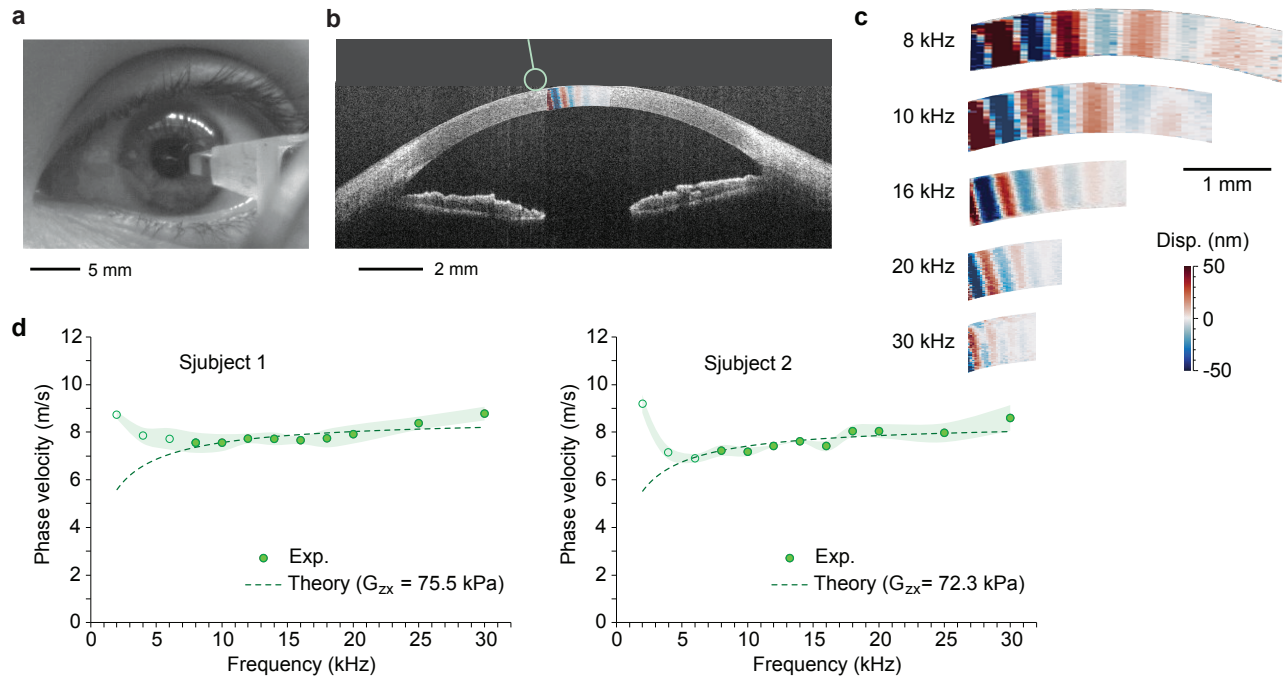

Figure 2. (a) Photograph showing the OCE measurement being performed on a volunteer. (b) A representative OCT image of the cornea. The circular symbol represents the probe location, and an overlaid OCE wave image measured at a frequency of 16 kHz is shown. (c) Wave images at various wave frequencies from 8 to 30 kHz, all captured at the same probe location. As the frequency increases, both the wavelength and propagation distance decrease. (d) Measured elastic wave speeds and calculated shear moduli obtained from two volunteers. The data points (circles) and shades represent the mean and standard deviation over three scan measurements. Low-frequency data below 6 kHz are susceptible to artifacts caused by wave interference. Dashed curves represent theoretical fits to the data above 6 kHz.

When $\sigma = 0$, $\bar{E}_{xx}$ and $\bar{G}_{zx}$ reduce to the intrinsic Young's modulus, $E_{xx}^{\sigma=0}$, and shear modulus, $G_{zx}^{\sigma=0}$, of the cornea in the stress-free condition [16].

We adopt the following strategy to determine the elastic moduli: First, we observe that the wave speed $v$ approaches approximately $0.96\alpha$ at high frequencies when the elastic wavelength is comparable to or smaller than the corneal thickness (Appendix Fig. A1). Hence, $\alpha$ is readily determined from high-frequency OCE data. Second, we estimate the ratio $\beta/\alpha$ from measured wave displacement profiles throughput the depth. Finally, $\gamma$ is determined from the relation $\gamma = \alpha - \sigma$, where $\sigma$ is calculated based on the IOP. Notably, $\bar{E}_{xx}/\bar{G}_{zx} = 2(\beta/\alpha + 1 - \sigma/\alpha)$. We will utilize this equation later.

In general, the magnitudes of $\alpha$, $\beta$, and $\gamma$ can vary within the cornea. Conducting a full three-dimensional analysis of these parameters is beyond the scope of this study. Instead, our analysis assumes that the three parameters remain constant within a small region of interest spanning a few wavelengths. We measure the transverse variation by measuring $\alpha$, $\beta$, and $\gamma$ at different locations from the central cornea to the sclera.

To quantitatively describe the wave motion, we employed the acoustoelastic solution derived from a constitutive model by Gasser et al. [35], originally developed for arterial walls. This model accounts for the anisotropic stiffness enhancement caused by collagen fibers while neglecting fiber dispersion in the central cornea. Subsequently, we derive a secular equation using the incremental dynamic theory [33, 34] to determine the dispersion relation and modal shape of the A0 wave. Within the range of validity of the constitutive model, we estimated the nonlinear variation of $\bar{G}_{zx}$ and $\bar{E}_{xx}$ as a function of IOP.

D. *Finite element analysis*

To simulate the cornea's mechanical behavior, we developed a finite element analysis (FEA) model using Abaqus/standard software (Abaqus 6.12, Dassault Systèmes). In the simulations of elastic wave propagation, we employed a plane strain model in an annulus geometry. The corneal curvature and thickness were obtained from OCT images and incorporated into the model. The aqueous humor was represented as an acoustic medium, initialized with a pressure matching the IOP. The arbitrary Lagrangian-Eulerian (ALE) adaptive mesh was employed to re-mesh the deformed acoustic medium during the initialization. Elastic waves were induced by applying a local surface pressure that mimics the contact probe.

To determine the contact stiffness, we employed an axisymmetric model for the cornea and a rigid spherical shell (radius 0.5 mm) to represent the PZT probe indenter. To account for the increase in intraocular pressure caused by indentation, the aqueous humor was modeled using the fluid-filled cavity feature available in Abaqus/standard with constant inner volume. The contact stiffness was derived from the slope of a force-displacement curve, allowing us to accurately capture the interaction between the cornea and the indenter during the simulation.

In the FEA models, we utilized the same constitutive material model [35] as employed in the analytical analysis. We ensured mesh convergence by verifying that the simulation results were independent of the mesh density.

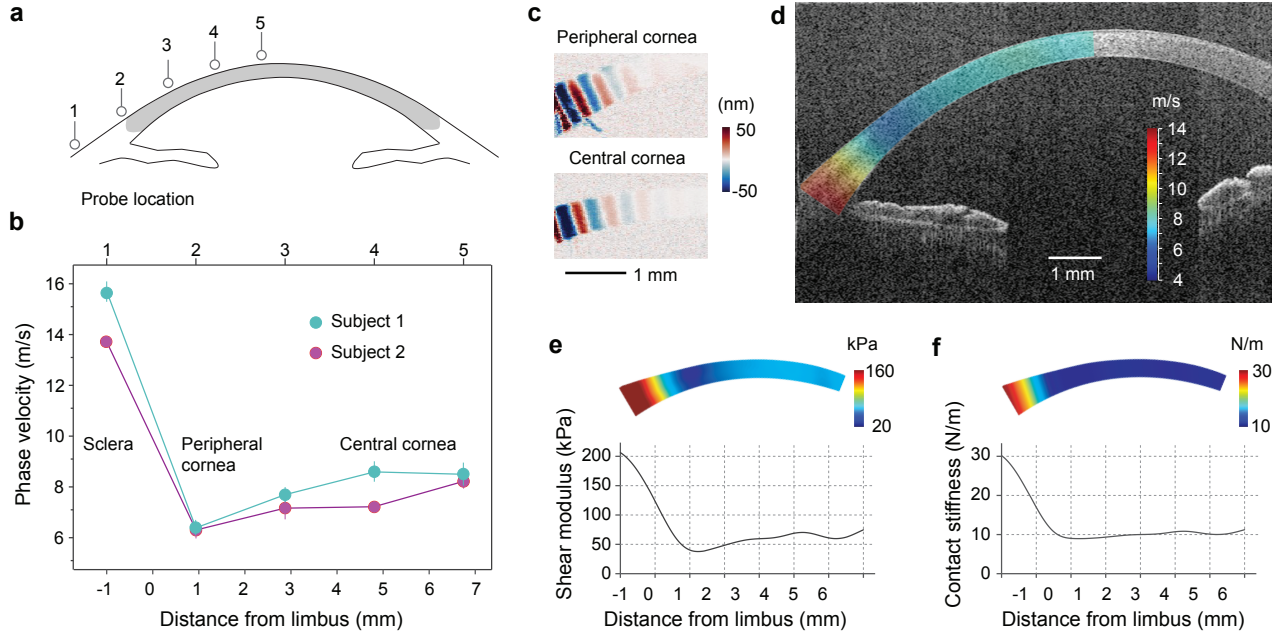

Figure 3. Spatial variations across the cornea. (a) Schematic showing the probe locations in five consecutive measurements. (b) Measured wave velocities from both subjects. The highest speed is observed in the sclera, while the lowest speed is found in the peripheral cornea. (c) Wave images of the peripheral and central corneas at 16 kHz. (d) Wave velocity map at 16 kHz overlaid on an OCT image (Subject 1). (e) Transverse profile of the measured shear modulus. (f) Contact stiffness profile calculated from the shear modulus map in (e). The slightly greater thickness in the peripheral cornea compensates for the lower shear modulus, result in relatively uniform bending stiffness rather uniform across the entire corneal region.

*E. Wave speed map reconstruction*

The wave speed map illustrating the shear modulus was reconstructed using the phase gradient algorithm [36, 37] with a window size about half wavelength of the elastic wave. At 16 kHz, the window size is ~ 0.25 mm. Within the window centered at $(x_0, z_0)$, orthogonal wave numbers $k_x$ and $k_z$ were calculated along the propagation direction $x$ and $z$. The 2-D wave speed map was calculated using

$$c = \frac{\omega_0}{\sqrt{k_x^2 + k_z^2}} \quad (4)$$

*F. Human study protocols*

This study was conducted following a protocol approved by Mass General Brigham Institutional Review Board (IRB). Two healthy male subjects, aged 30 and 32 years old, were recruited, both with moderate myopia (−3 diopters) and a nominal IOP of approximately 15 mmHg. Written informed consent was obtained from each subject after providing a detailed explanation of the study's nature and possible consequences. Only the left eyes of the subjects were scanned. Prior to OCE measurements, proparacaine ophthalmic drops were applied as a topical anesthetic to the left eye.

To determine the wave speed dispersion at the central cornea, we positioned the probe near the cornea's center (refer to Fig. 2a) and generated elastic waves at various frequencies ranging from 2 to 30 kHz. with a frequency increment of 2 or 5 kHz. To investigate the transverse variation of corneal stiffness, we maintained the frequency at 16 kHz and successively moved the probe to five locations along the left eye, ranging from the temporal sclera near the limbus to the central cornea, with a step size of approximately 2 mm. The wave velocity at each location was computed.

III. RESULTS

*A. Shear modulus of the cornea in vivo*

Figures 2b and 2c depict a representative OCT image of Subject 1 and the corresponding OCE wave images acquired at different frequencies. The wave speeds as a function of frequency for both subjects are presented in Fig. 2d. Below 6 kHz, the speed measurements are erroneous due to wave reflections from the corneal boundary. At 16 kHz, the elastic wavelength becomes shorter than the corneal thickness while generating sufficient vibrational amplitudes for reliable speed measurement. We employed curve fitting based on the acoustoelastic theory, incorporating various morphological and mechanical parameters reported in Refs. [38, 39], with the shear modulus $\bar{G}_{zx}$ as the sole fitting parameter. By obtaining the best-fit curves, we derived in-situ shear modulus values of 75.5 ± 5.0 kPa for Subject 1 and 72.3 ± 4.6 kPa for Subject 2 (± represents 95% confidence level of the curve fitting). These values align with our previous mean value of 72 kPa, obtained from 12 subjects, exhibiting an interpersonal variability of 14 kPa, through simpler Kevin-Voight model fitting [16].

Using the constitutive model, we estimated the "intrinsic" shear modulus of the cornea under zero tension and zero strain, which theoretically could be measured from excited corneal tissues. These estimated the mean intrinsic shear modulus values ($G_{zx}^{\sigma=0}$) were 64.2 kPa from Subject 1 and 62.5 kPa from Subject 2. As expected, these values were lower than the shear

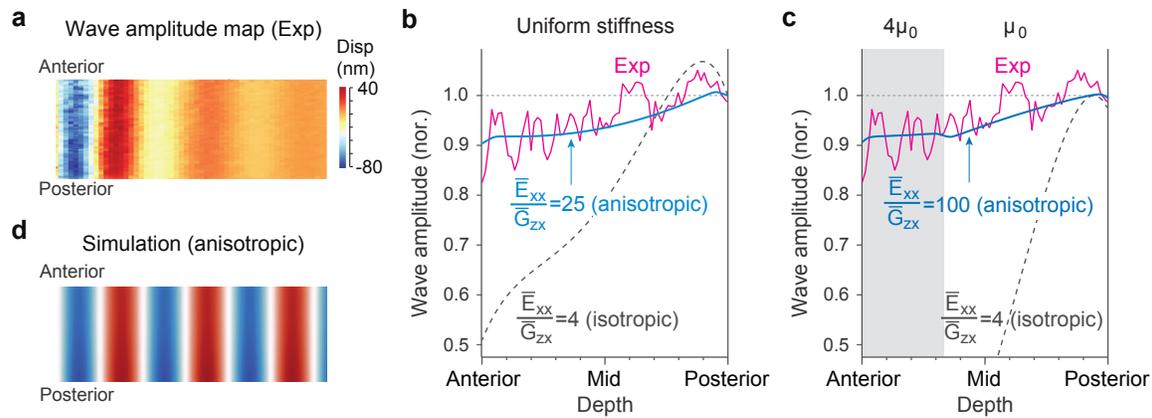

Figure 4. Anisotropic property of the corneal stroma. (a) Wave map measured at 16 kHz in the central cornea. (b) Normalized displacement profile (magenta) across the depth, along with a best-fit simulation result for a uniform, anisotropic tissue under tension (blue), and a theoretical curve for an isotropic material (dashed curve). (c) Same as (b), but for a nonuniform tissue with depth-dependent stiffness (the top 1/3 region has 4 times higher shear modulus than the bottom 2/3 thickness). (d) Simulated wave profile assuming no viscous damping (corresponding to the blue curve in (b)).

modulus of the cornea *in situ* and similar to the value reported previously [40].

### B. Spatial variation of shear modulus in the cornea

Figure 3 illustrates the wave speeds measured at different locations within the eye. The wave speeds in the sclera are 15.7 ± 0.45 m/s for Subject 1 and 13.7 ± 0.4 m/s for Subject 2, corresponding to approximately 2-fold values compared to the average speeds observed in the cornea and yielding 4-fold higher shear moduli. Notably, the peripheral region in the cornea, located approximately 1 mm away from the limbus, exhibits the lowest wave speeds (6.4 ± 0.2 m/s for Subject 1 and 6.3 ± 0.15 m/s for Subject 2). This corresponds to in-situ shear moduli ($\bar{G}_{zx}$) of 41 kPa and 40 kPa, respectively, which is nearly half the value (~74 kPa) observed in the central cornea. Figure 3c provides a comparison between wave images captured in the peripheral and central regions at 16 kHz, clearly demonstrating the shorter wavelength in the peripheral cornea. Additionally, Fig. 3d presents a wave speed map for Subject 1, reconstructed using a phase gradient algorithm (see Method). This speed map reveals striking gradients in corneal mechanics, with a spatial resolution of approximately 0.5 mm (the wavelength at 16 kHz).

The lower speeds observed in the peripheral cornea may be attributed to two factors. Firstly, the peripheral cornea is about 20% thicker than the central cornea (0.64 mm versus 0.53 mm) [38]. Consequently, at a given IOP, the greater thickness results in a lower tensile stress in the peripheral cornea [as per Eq. (1)], leading to a decrease in wave speed via the acoustoelastic effect. Secondly, the peripheral cornea may possess an intrinsic softness compared to the central cornea. To verify this, we conducted finite element simulations. Assuming the cornea had uniform intrinsic material properties, the wave speed in the thicker periphery was only 4% lower than in the central cornea. However, the significant variation in wave speed observed could only be replicated when the shear modulus in the peripheral cornea was estimated to be 64% of that in the central cornea. Previous *ex vivo* studies on human corneas have also reported similar gradients. For instance, an ultrasound study demonstrated that the shear modulus of the anterior region of the cornea varied from 1.4 kPa in the center to 0.6 kPa in the periphery [39]. Although their modulus values are considerably lower, likely influenced by swelling in cadaver tissues, the ratio of difference appears comparable. Figure 3e exhibits a shear modulus profile derived from the speed gradient map using the acoustoelastic theory. The shear modulus in the limbus is higher than 100 kPa.

Next, we evaluated the contact stiffness, which characterizes the cornea's response to local surface indentation. This property is extrinsic and influenced by both intrinsic elastic modulus and morphology. We conducted finite element simulations to calculate the displacement slope as a function of indentation load using a rigid sphere (radius 0.5 mm) (see Appendix Fig. A2). As depicted in Fig. 3f, the contact stiffness remains relatively uniform throughout the cornea, primarily due to the contribution of thickness ($h$) to bending stiffness (proportional to $h^3$) during indention.

### C. Anisotropy of tensile and shear moduli

According to the standard acoustic theory for a uniform plate made of isotropic material, the wave profile of the A0 mode would deviate significantly from a uniform shear-like motion. At low frequencies where the wavelength exceeds the thickness, the A0 mode exhibits flexural (bending) motion, while at high frequencies, the wave becomes increasingly localized at the surface. However, the *in vivo* wave images we measured displayed nearly bulk shear wave-like motion (see Fig. 2c and Fig. 4a). To understand this phenomenon, we conducted finite element analysis based on the acoustoelastic theory, confirming that the distinctive wave profile arises from the anisotropy or the significance difference between $\bar{G}_{zx}$ and $\bar{E}_{xx}$. In our model, for a given dimensionless wave number $kh$, the displacement profile is primarily influenced by the $\beta/\alpha$ ratio but is less sensitive to the $\gamma/\alpha$ ratio (see Appendix Fig. A3).

Figure 4b shows a typical depth profile measured at 16 kHz in the central cornea. In the finite element simulation, we initially assumed spatially uniform elastic moduli and varied the $\beta/\alpha$ ratio to fit the experimental data. Figure 4b showcases the wave depth profiles for two representative cases: $\bar{E}_{xx}/\bar{G}_{zx} = 4$ (3 for nonplanar elastic waves), corresponding to isotropic materials without tension (represented by dashed curve), and the best-fit case with $\beta/\alpha \approx 12$. Using $\gamma/\alpha \approx 0.8$ from the IOP, we obtained $\bar{E}_{xx}/\bar{G}_{zx} = (2\beta + 2\gamma)/\alpha \approx 25$. With $\bar{G}_{zx} \approx 74$ kPa, we estimate $\bar{E}_{xx}$ to be approximately 1.8 MPa.

Previous mechanical testing of *ex vivo* corneal tissues (typically without external tension) revealed that the anterior stroma is significantly stiffer than the mid and posterior stroma [25, 26]. Rheometry measurements have reported a shear modulus of $7.7 \pm 6.3$ kPa in anterior layers, 3-5 times higher than the shear modulus of $2.0 \pm 0.45$ kPa in middle layers and $1.3 \pm 1.0$ kPa in the posterior layers [41]. Such a downward gradient of modulus should result in a more localized wave profile in the lower layers. To counteract the tendency, even greater anisotropy is required to produce shear-like waves. We conducted finite element simulation for a stepwise case, where the top 1/3 region had a shear modulus four times higher than the bottom 2/3 region. Notably, the best fit for the nonuniform cornea was obtained with $\beta/\alpha \approx 50$ (Fig. 4C), corresponding to $\bar{E}_{xx}/\bar{G}_{zx} \approx 100$ and $\bar{E}_{xx} \approx 7.2$ MPa. These estimated values are comparable to the tensile moduli of 0.8-2.2 MPa previously reported in *ex vivo* samples through quasi-static stress-strain testing [7].

## IV. Discussion

The analysis of acoustoelastic data in our study provided a comprehensive understanding of the corneal wave characteristics by considering the effects of collagen fiber arrangement and IOP-induced tension. By utilizing the derived acoustoelastic solution in conjunction with the incremental dynamic theory, we achieved accurate estimation of wave dispersion and modal shape for the A0 wave, while also considering the influence of IOP on the nonlinear variation of the out-of-plane shear modulus and in-plane tensile modulus. This approach allowed for a detailed quantitative description of corneal wave behavior and its relationship with IOP.

In our human pilot study, we measured the shear modulus of the central cornea to be approximately 74 kPa (at 16 kHz) under normal IOP. According to the constitutive material model, the intrinsic shear modulus without corneal tension (as observed in excited corneas) was estimated to be 63 kPa, with the tension-induced stiffening of collagen fibrils in the stroma accounting for an 11 kPa difference.

The high-frequency OCE technique utilized in this study provided excellent spatial resolution, surpassing 1 mm. We observed a gradual decrease in tissue stiffness from the central cornea to a minimum shear modulus of 41 kPa in the peripheral cornea, approximately 1 mm from the limbus. The underlying structural and physiological reasons for the low stiffness in the corneal periphery remain unclear. The shear modulus of the limbus was higher than the cornea but lower than the sclera. Further investigation into the biomechanics of the limbal region, with higher spatial resolution achieved using frequencies even higher than 30 kHz, would be valuable. Conversely, the increased thickness in the peripheral cornea compensates for the reduced shear modulus, resulting in contact stiffness that remains relatively uniform throughout the cornea. This finding has implications for applanation or air-puff tonometry, where significant corneal deformation occurs. Thus, the peripheral region is softer but not weaker from a mechanical standpoint.

Through the acoustoelastic analysis of the measured wave profiles, we estimated the tensile (Young's) modulus of the central cornea to be approximately 7 MPa when assuming that the anterior cornea is four times stiffer than the mid and posterior regions. This value is subject to some uncertainty within a few MPa, depending on the exact stiffness profile along the depth. A more direct way to measuring tensile stiffness may involve employing a symmetric (S0) Lamb wave in addition to the A0 mode, which is currently under development in our laboratory.

## V. Conclusion

In this work, we developed a high-frequency OCE technique that allows for measuring *in situ* corneal elastic moduli with high precision (<7%) and high spatial resolution (< 1mm). Using the proposed method, we observed remarkable stiffness gradients in human cornea *in vivo* for the first time. By incorporating an anisotropic, nonlinear constitutive model and utilizing the acoustoelastic theory, we further quantified the tensile and shear moduli in the cornea under IOP-induced tension. Our technique has significant potential for clinical applications, by enabling physicians to obtain patient-specific *in vivo* data that accurately reflect corneal properties *in situ*, considering physiological IOP, tension, and hydration levels.

## Appendix

### A. Acoustoelastic model for corneal OCE

Here we derive the mechanical model for corneal OCE. To incorporate mechanical anisotropy and mechanical loading, we build our model on the acoustoelastic theory. Readers can refer to Ref [42] for details of the theory.

**Wave equation.** According to the acoustoelastic theory, the wave equation for small-amplitude plane elastic wave in a uniformly prestressed solid reads [34]

$$\alpha \frac{\partial^4 \psi}{\partial x_1^4} + 2\beta \frac{\partial^4 \psi}{\partial x_1^2 \partial x_3^2} + \gamma \frac{\partial^4 \psi}{\partial x_3^4} = \rho \left( \frac{\partial^4 \psi}{\partial x_1^2 \partial t^2} + \frac{\partial^4 \psi}{\partial x_3^2 \partial t^2} \right), \quad (A1)$$

where we have used $x_1$ and $x_3$ to denote the coordinates that correspond to $x$ and $z$ in the main text, respectively. The stream function $\psi$ relates to displacement components $u_1$ and $u_3$ via the relation of $u_1 = \partial \psi / \partial x_3$ and $u_3 = -\partial \psi / \partial x_1$. It satisfies $\partial u_1 / \partial x_1 + \partial u_3 / \partial x_3 = 0$, the constraint equation for material incompressibility. $\rho$ and $t$ denote the density and time, respectively. The coefficients $\alpha$, $\beta$, and $\gamma$ are determined by the constitutive law and the stretch ratio $\lambda$,

$$\alpha = \mathcal{A}_{1313}^0, \quad (A2)$$

$$2\beta = \mathcal{A}^0_{1111} + \mathcal{A}^0_{3333} - 2\mathcal{A}^0_{1133} - 2\mathcal{A}^0_{1331},$$

$$\gamma = \mathcal{A}^0_{3131},$$

where the fourth-order tensor, $\mathcal{A}^0_{ijkl}$, is the Eulerian elasticity tensor and is defined as

$$\mathcal{A}^0_{ijkl} = F_{iI}F_{kJ}\frac{\partial^2 W}{\partial F_{jI}\partial F_{lJ}}, \quad i,j,k,l,I,J \in \{1,2,3\}, \tag{A3}$$

where $F_{ij}$ is the deformation gradient tensor (F), and $W$ is the strain energy function. In Eq. (A3) the Einstein's summation convention is used. For the Holzapfel-Gasser-Ogden (HGO) model, the strain energy is

$$W = \frac{\mu}{2}(I_1 - 3) + \frac{k_1}{k_2}\sum_{i=1}^2\left\{e^{k_2[\kappa(I_1-3)+(1-3\kappa)(I_{4i}-1)]^2} - 1\right\} \tag{A4}$$

where $\mu$, $k_1$, $k_2$ and $\kappa$ are constitutive parameters. $\mu$ denotes the shear modulus in stress-free state. The dimension of $k_1$ is the same as $\mu$, whereas $k_2$ is a dimensionless parameter which determines the nonlinear hardening effect of the collagen fibrils when being stretched. $\kappa = 0$ if the collagen fibrils are ideally aligned [43], which is applicable for the cornea. $I_1 = \text{tr}(F^T F)$. $I_{41}$ and $I_{42}$ are two invariants related to two families of collagen fibers. Following the coordinates defined in Fig. 1b, the axes of the collagen fibers of the cornea, denoted by unit vectors M and M′, are aligned with ($x_1$ and $x_2$, i.e., M = $(1,0,0)^T$ and M′ = $(0,1,0)^T$. Then $I_{41}$ and $I_{42}$ can be determined by M and M′ [35]:

$$I_{41} = (FM) \cdot (FM), \quad I_{42} = (FM') \cdot (FM'). \tag{A5}$$

Then, the coefficients $\alpha$, $\beta$, and $\gamma$ of the wave equation can be obtained by inserting Eqs. (A3) and (A4) into Eq. (A2)

$$\alpha = \lambda^2\left\{\mu + 2k_1(\lambda^2-1)e^{[k_2(\lambda^2-1)^2]}\right\},$$
$$\gamma = \mu\lambda^{-4}, \tag{A6}$$
$$2\beta = \alpha + \gamma + 4k_1\lambda^4[2k_2(\lambda^2-1)^2 + 1]e^{[k_2(\lambda^2-1)^2]}.$$

In the absence of prestress (i.e., $\lambda = 1$)

$$\alpha = \mu, \beta = \mu + 2k_1, \gamma = \mu. \tag{A7}$$

**Dispersion relation.** We now consider the guided wave motion in the cornea. The two sides of the cornea are interfaced with the air and aqueous humor, respectively. The aqueous humor is modeled as a semi-infinite fluid layer and the wave equation is

$$\nabla^2 \chi = \frac{\rho^f}{v}\chi_{,tt}, \tag{A8}$$

where $v$ (2.2 GPa) and $\rho^f$ (1,000 kg/m³) denote the bulk modulus and density of the fluid, respectively. $\chi$ is a potential function related to the displacement of the fluid (denoted by $\mathbf{u}^f$) through of $u_1^f = \chi_{,1}$ and $u_2^f = \chi_{,2}$. The pressure of the fluid, denoted by $p^*$, is determined by

$$p^* = -v\nabla \cdot \mathbf{u}^f. \tag{A9}$$

At the interface between cornea and aqueous humor ($x_3 = 0$), the following interfacial conditions apply

$$u_3 = u_3^f, \quad -\gamma\psi_{,11} + \gamma\psi_{,33} = 0,$$
$$\rho\psi_{,3tt} - (2\beta+\gamma)\psi_{,113} - \gamma\psi_{,333} = -p_{,1}^*. \tag{A10}$$

At the surface of cornea ($x_3 = h$), the stress-free boundary conditions require

$$-\gamma\psi_{,11} + \gamma\psi_{,22} = 0,$$
$$\rho\psi_{,2tt} - (2\beta+\gamma)\psi_{,112} - \gamma\psi_{,222} = 0. \tag{A11}$$

More details on the derivations of the boundary conditions can be found in previous studies [33, 44].

We seek the plane wave solutions for $\psi(x_1, x_3, t)$ and $\chi(x_1, x_3, t)$, i.e.,

$$\begin{cases} \chi(x_1, x_3, t) = \chi_0(x_3)e^{\iota k(x_1-ct)} \\ \psi(x_1, x_3, t) = \psi_0(x_3)e^{\iota k(x_1-ct)}, \end{cases} \tag{A12}$$

where $\iota = \sqrt{-1}$, $k$ is the wavenumber, and $c$ is the phase velocity. Inserting Eq. (A12) into Eqs. (A1) and (A8), we get

$$\begin{cases} \chi = Ae^{-\xi k x_3}e^{\iota k(x_1-ct)} \\ \psi = [B_1\cosh(s_1 k x_3) + B_2\sinh(s_1 k x_3) \\ +B_3\cosh(s_2 k x_3) + B_4\sinh(s_2 k x_3)]e^{\iota k(x_1-ct)}. \end{cases} \tag{A13}$$

The parameters $s_1$, $s_2$ and $\xi$ are determined by

$$\gamma s^4 - (2\beta - \rho c^2)s^2 + (\alpha - \rho c^2) = 0, \tag{A14}$$

and

$$\xi^2 - 1 = -c^2\rho^f/v. \tag{A15}$$

Substituting $\psi(x_1, x_2, t)$ and $\chi(x_1, x_2, t)$ into Eqs. (A10) and (A11), we get

$$M_{5\times 5} \cdot [B_1, B_2, B_3, B_4, A]^T = 0, \tag{A16}$$

where the nonzero components of the 5x5 matrix $M$ are

$$\begin{aligned}
&M_{11} = s_1^2 + 1, M_{13} = s_2^2 + 1,\\
&M_{22} = \gamma s_1(s_2^2 + 1),\\
&M_{24} = \gamma s_2(s_1^2 + 1), M_{25} = \iota\rho^f c^2,\\
&M_{31} = 1, M_{33} = 1, M_{35} = -\iota\xi,\\
&M_{41} = (s_1^2 + 1)\cosh(s_1 kh),\\
&M_{42} = (s_1^2 + 1)\sinh(s_1 kh),\\
&M_{43} = (s_2^2 + 1)\cosh(s_2 kh),\\
&M_{44} = (s_2^2 + 1)\sinh(s_2 kh),\\
&M_{51} = s_1(s_2^2 + 1)\sinh(s_1 kh),\\
&M_{52} = s_1(s_2^2 + 1)\cosh(s_1 kh),\\
&M_{53} = s_2(s_1^2 + 1)\sinh(s_2 kh),\\
&M_{54} = s_2(s_1^2 + 1)\cosh(s_1 kh).
\end{aligned} \tag{A17}$$

In this derivation, we have utilized the identity

$$2\beta - \rho c^2 = \gamma(s_1^2 + s_2^2), \tag{A18}$$

which can be obtained from Eq. (A14).

The dispersion relation can be obtained by solving equation

$$\det(M_{5\times 5}) = 0, \quad (A19)$$

The nontrivial solution of $[B_1, B_2, B_3, B_4, A]^T$ gives the modal shape ($0 \leq x_3 \leq h$). The vertical displacement is given by:

$$u_3 = \left|\left(\frac{2\pi f}{v}\right) \times [B_1 \cosh(s_1 k x_3) + B_2 \sinh(s_1 k x_3) + B_3 \cosh(s_2 k x_3) + B_4 \sinh(s_2 k x_3)]\right|. \quad (A20)$$

### B. Relating IOP to stretch ratio $\lambda$

The in-plane stress within the cornea that balances the intraocular pressure can be determined by the Young-Laplace equation: $\sigma = IOP \times R/(2h)$. Here we show how to relate the stress to the stretch ratio $\lambda$, which is involved in the dispersion relation.

The strain energy function relates the deformation to Cauchy stress by

$$\sigma_{ij} = F_{il} \partial W/\partial F_{jl} - \bar{p}\delta_{ij}, \quad (A21)$$

where $\sigma_{11} = \sigma_{22} = \sigma$, $\bar{p}$ is a Lagrange multiplier for material incompressibility and $\delta_{ij}$ is the Kronecker delta. For cornea the out-of-plane stress $\sigma_{33}$ is negligible in comparison to the in-plane stress $\sigma$. So we can get

$$\sigma = \mu(\lambda^2 - \lambda^{-4}) + 2k_1\lambda^2(\lambda^2 - 1)e^{[k_2(\lambda^2-1)^2]}. \quad (A22)$$

The stretch ratio $\lambda$ can be obtained by solving the nonlinear equation

$$\mu(\lambda^2 - \lambda^{-4}) + 2k_1\lambda^2(\lambda^2 - 1)e^{[k_2(\lambda^2-1)^2]} = IOP \times R/(2h). \quad (A23)$$

Equation (A23) relates IOP to the deformation $\lambda$. Notably, it is straightforward to check

$$\sigma = \alpha - \gamma. \quad (A24)$$

For normal human subjects with typical $\sigma$ of 15 kPa and $\alpha$ of 70 kPa, we obtain $\gamma/\alpha \approx 0.8$.

### C. Lamb waves in cornea

Appendix Fig. A1 shows representative dispersion relations of the fundamental Lamb waves (A0 and S0) in cornea obtained from our theoretical model. At zero frequency, $f = 0$, the phase velocity of the S0 mode is $\sqrt{(2\beta + 2\gamma)/\rho}$. When $f \to +\infty$, the phase velocities of A0 and S0 become the Scholte and Rayleigh surface wave speeds, respectively, which are primarily determined by $\sqrt{\alpha/\rho}$. The two phase velocities, in the absence of prestress, are related to plane-strain Young's modulus ($E_{xx} = E_{yy}$; here the subscript from $E_{ii}^{\sigma=0}$ has been dropped for simplicity) and shear modulus ($G_{zx}$) by $\sqrt{E_{xx}/\rho}$ and $\sqrt{G_{zx}/\rho}$, indicating $(2\beta + 2\gamma)$ and $\alpha$ reduce to the Young's modulus and shear modulus, respectively. Here $E_{xx}$, $E_{yy}$, and $G_{zx}$ are components of the stiffness matrix of the intrinsic corneal tissue in the stress-free condition.

Inspired by this observation, we can use $(2\beta + 2\gamma)$ and $\alpha$ to characterize the *in-situ* tensile and shear stiffness (denoted by $\bar{E}_{xx}$ and $\bar{G}_{zx}$) of the cornea under tension. According to Figure A1, the dispersion relation of A0 is sensitive to $\bar{G}_{zx}$, and the dispersion relation of S0 in the low frequency regime is sensitive to $\bar{E}_{xx}$, which suggests we can interrogate shear and tensile stiffness of cornea by probing A0 and S0 Lamb waves, respectively.

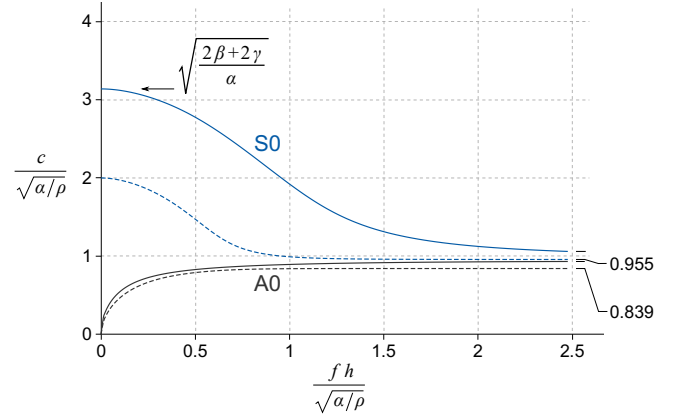

Figure A1. Dimensionless dispersion relations of the A0 and S0 mode Lamb waves. Solid lines, $\beta/\alpha = 4$, $\gamma/\alpha = 0.92$. Dashed lines, isotropic materials without prestress ($\alpha = \beta = \gamma$). The phase velocity of S0 at $f = 0$ is $\sqrt{(2\beta + 2\gamma)/\rho}$, governed by the tensile stiffness $(2\beta + 2\gamma)$. In the high frequency regime, the phase velocities of A0 and S0 reach plateaus, which are Scholte (solid-fluid interface) wave and Rayleigh wave speed, respectively.

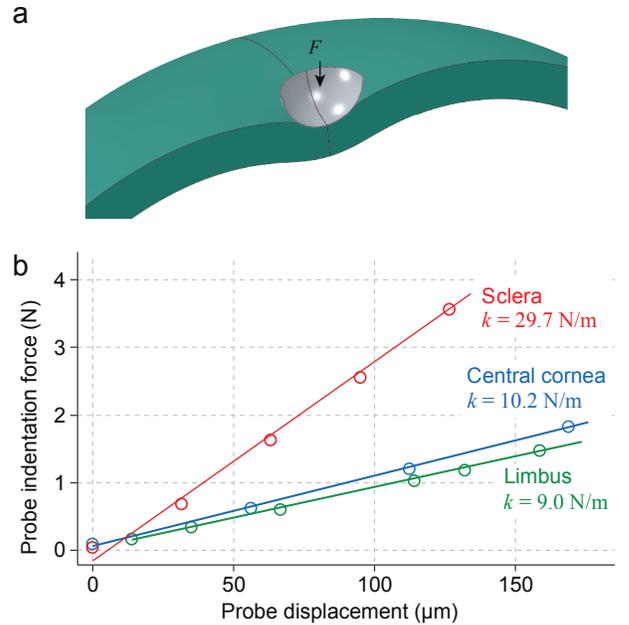

Figure A2. Finite element analysis of the local contact stiffness for the cornea. (a) a presentative corneal deformation. (b) The load-force curve obtained from different locations.

Here:

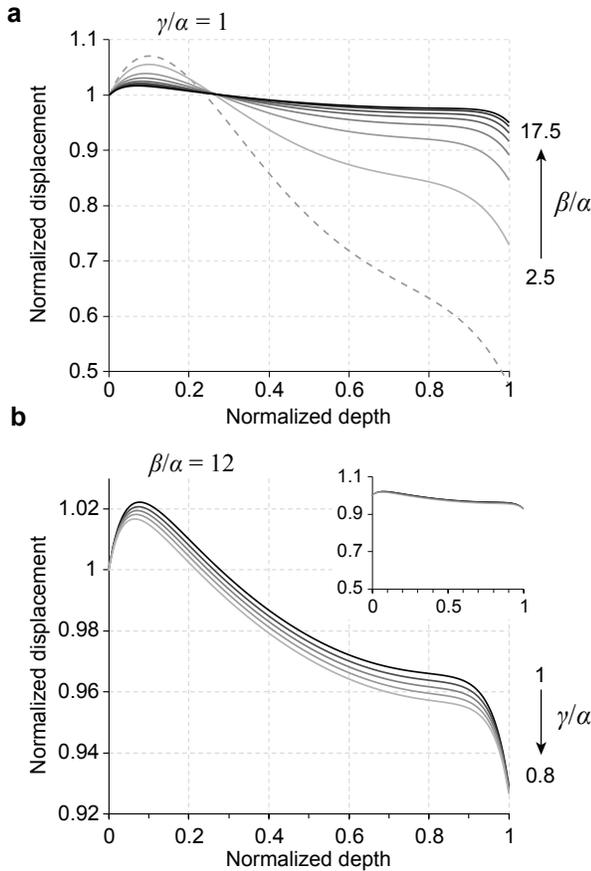

Figure A3. Modal shapes for $kh = 1.5\pi$. (A) Variation of the modal shape with the ratio $\beta/\alpha$ when $\gamma/\alpha = 1$. (B) Variations of the modal shape with the ratio $\gamma/\alpha$ when $\beta/\alpha = 12$.